\documentstyle[revtex,preprint]{aps}
\newcommand{\Frac}[2]{\frac{{\displaystyle #1}}{{\displaystyle #2}}}

\begin{document}\draft

\begin{title} Attractive Forces Between Electrons in QED$_{3}$ \end{title}

\author{H. O. Girotti \cite{ast}, M. Gomes, J. L. deLyra, R. S. Mendes and J.
R. S. Nascimento}

\begin{instit} Instituto de F\'{\i}sica, Universidade de S\~ao Paulo \\ Caixa
Postal 20516, 01498 - S\~ao Paulo, SP, Brazil. \end{instit}

\author{A. J. da Silva \cite{loa}}

\begin{instit} Center for Theoretical Physics, Massachusetts Institute of
Technology \\ Cambridge, MA 02139 \end{instit}

\begin{center}
 To appear in Phys. Rev. Lett. (november/1992)
\end{center}
\begin{abstract}

Vacuum polarization effects are non-perturbatively incorporated into the photon
propagator to eliminate the severe infrared problems characteristic of QED$_3$.
The theory is thus rephrased in terms of a massive vector boson whose mass is
$e^2/(8\pi)$. Subsequently, it is shown that electron-electron bound states are
possible in QED$_3$.

\end{abstract}

\pacs{PACS: 11.10.St, 11.15.Tk}

\narrowtext

Recently \cite{Gi}, the effective non-relativistic potential

\begin{eqnarray}\label{at1}V(\vec r)&=&\frac{e}{2\pi}\left(1-
\frac{\theta}{m}\right)K_{0}\left(|\theta|r\right)\nonumber\\&&
-\frac{e}{\pi\theta}\frac{1}{mr^{2}}\left[1-|\theta|r K_{1}\left(|
\theta|r\right)\right]L,\end{eqnarray}

\noindent describing the fermion-fermion interaction in the
Maxwell-Chern-Simons (MCS) theory \cite{Ja}, was derived in the lowest
perturbative order (the notation in Eq. (\ref{at1}) is explained in Ref.
\cite{f1})\cite{f5}. The aim in Ref.~\cite{Gi} was to determine
whether the potential (\ref{at1}) could bind a pair of identical fermions. For
positive values of $\theta$, a numerical solution of the Schr\"odinger equation
confirmed the existence of a bound state for $e^2/(\pi\theta)=500$,
$m/\theta=10^5$ and $l=1$ . Further numerical analysis
indicated that all identical-fermion bound states are located in the
region $e^2/(\pi\theta)>1$.

Thus, we were naturally led to study the limit $\theta\rightarrow 0$ where the
MCS theory degenerates into QED$_3$. Power counting indicates that QED$_3$ is
plagued with infrared singularities whose degree of divergence grows with the
order of perturbation. Moreover, when QED$_3$ is treated non-perturbatively by
means of the Bloch-Nordsieck (BN) approximation \cite{Bl,Bo,Sz}, one finds
 for the two-point fermionic Green's function of momentum $p$ the
expression

\FL\begin{eqnarray}\lefteqn{G(p;m_R)=-\frac{i}{|Q|}}&&\nonumber\\
&\times&\int_{0}^{\infty}dv\exp\left\{\epsilon(Q)v-\frac{e^2}{4\pi|Q|}v
\left[C+\ln\left(\frac{M}{|Q|}v\right)\right]\right\},\end{eqnarray}

\noindent where $Q\equiv u\cdot p-m$, $u$ is a time-like vector ($u^2=1$)
replacing
the gamma matrices in the BN scheme, $\epsilon(Q)$ is the sign function, $M$ is
a subtraction point and $C$ is the Euler constant. Clearly, $G$ is well
defined for generic values of $Q$ but develops an essential singularity at
$Q=0$. The fact that $G$ does not behave as a simple pole at $Q=0$ signalizes
the presence of infrared singularities.

As is known \cite{Bl}, the BN approximation eliminates all vacuum
polarization diagrams \cite{f2}. In this work we start by demonstrating that
when the vacuum polarization effects are non-perturbatively incorporated into
the photon propagator, the infrared structure of QED$_3$ changes drastically
and all inconsistencies disappear. Essentially, the theory is reformulated in
terms of a massive vector boson whose mass is $|\theta |=e^2/(8\pi)$, the
dynamically induced Chern-Simons term \cite{Ja,Re} being at the root of this
mechanism. It turns out then that the effective electron-electron low-energy
potential arising from QED$_3$ can be read off directly from (\ref{at1}) after
the replacement $\theta\rightarrow-e^2/(8\pi)$, namely,

\begin{eqnarray}\label{at3}eV^{\rm QED_{3}}(\vec r)&=&\frac{e^2}{2\pi}\left(
1+\frac{e^2}{8\pi m}\right)K_{0}\left(\frac{e^2}{8\pi}r\right)\nonumber
\\&&+\frac{8}{mr^{2}}\left[1-\frac{e^2}{8\pi}r K_{1}\left(\frac{e^2}{8\pi}
r\right)\right]L.\end{eqnarray}

\noindent The terms proportional to $K_{0}$ in (\ref{at3}) are now both
repulsive. The term $8L/(mr^2)$ becomes attractive (repulsive) for negative
(positive) eigenvalues of $L$, while the term proportional to $K_{1}$ acts in
the opposite way. We conclude the paper by showing that electron-electron bound
states are also possible in QED$_3$.

Our starting point is the QED$_3$ Lagrangian density \cite{f3}

\begin{eqnarray}\label{at4}{\cal L}&=&-\frac{1}{4}F_{\mu\nu}F^{\mu
\nu}-\frac{1}{2\lambda}(\partial_{\mu}A^{\mu})(\partial_{\nu}A^{\nu})+
\frac{i}{2}\bar{\psi}\gamma^{\mu}\partial_{\mu}\psi\nonumber\\&&
\mbox{}-\frac{i}{2}(\partial_{\mu}\bar{\psi})\gamma^{\mu}\psi+e
\bar{\psi}\gamma^{\mu}A_{\mu}\psi-\bar m\bar{\psi}\psi,\end{eqnarray}

\noindent describing the coupling of charged fermions ($\bar\psi,\psi$) of mass
$m =|\bar m|$ and charge $e$ to the gauge field potential $A^\mu$. In
principle, $\bar m$ can be either positive or negative, but we shall analyze
here the case $\bar m>0$. Neither parity nor time-reversal are, separately,
symmetries of the model.

We concentrate on the lowest-order graph contributing to the vacuum
polarization tensor $\Pi^{\rho\sigma}(q)$. Since we are interested in the
quantum corrections to a non-relativistic potential, we shall retain only those
terms of zero and first order in $q$. Gauge invariance alone ensures that
$\Pi^{\rho\sigma}(0)=0$. As for the first-order contribution, which gives
origin to the induced Chern-Simons term \cite{Ja,Re}, one finds

\begin{equation}
\Pi_{\rho\sigma}^{(1)}(q)=-i\frac{e^2}{8\pi}\epsilon_{\rho\sigma\mu}q^{\mu}.
\label{5}
\end{equation}
\noindent We emphasize that $\Pi_{\rho\sigma}^{(1)}(q)$ is ultraviolet-finite,
and that, therefore, no regularization is needed for its computation. At this
point a word of caution is necessary. If one adopts the point of view
of keeping ultraviolet divergences under controll by regularizing the entire
theory, the result quoted in this last equation is only true if a parity--time
reversal invariant regularization is used.

Because of the severe infrared singularities, the standard perturbative series
fails to exist in QED$_3$. On the other hand, one knows that when  a
Chern-Simons term is added to the free part of the QED$_3$
Lagrangian (the MCS theory), a topological
mass for the vector field is generated, freeing the theory from the infrared
divergences. In view of this, we modify the photon propagator by resumming
the geometric series resulting from the iteration of $\Pi^{(1)}_{\rho\sigma}$
(see Fig.~\ref{fi1}). As far as the nonrelativistic approximation is
concerned, this is equivalent to the introduction into the free Lagrangian of
the Chern-Simons term

\begin{equation}{\cal L}_{CS}=-\frac{e^2}{32\pi}\epsilon^{\mu\nu\rho}F_{\mu\nu}
A_{\rho}.\end{equation}

\noindent Hence, the induced Chern-Simons coefficient is, as previously stated,

\begin{equation}\label{at7}\theta_{in}=-\frac{e^2}{8\pi}.\end{equation}

\noindent From (\ref{at1}) and (\ref{at7}) it then follows that the QED$_3$
effective electron-electron low-energy potential is, in fact, that given by
(\ref{at3}).

Before investigating whether the potential (\ref{at3}) can sustain
electron-electron bound states, we want to show that the above proposed
solution
for QED$_3$ is consistent, in the sense that all remaining
contributions to $V^{\rm QED_{3}}$ are, up to some power of $\ln(e^2/m)$, of
order $e^2/m$ or higher with respect to (\ref{at3}) and, therefore, vanish as
$e^2/m\rightarrow 0$. To see how this comes about, we compute the vertex
correction $\Lambda^\mu$ to $V^{\rm QED_{3}}$ arising from the diagram in which
only one massive vector particle is exchanged (see Fig.~\ref{fi2}) \cite{f7}.
After the
replacement $\theta\rightarrow\theta_{in}$, the massive vector field propagator
can be read off directly from Eq.~(4) of Ref.~\cite{Gi},

\begin{eqnarray}D_{\mu\nu}(k)=&&\frac{-i}{k^{2}-\theta^{2}_{in}}\left
(P_{\mu\nu}\,\,-\,\,i\theta_{in}\epsilon_{\mu\nu\rho}\frac
{k^{\rho}}{k^{2}}\right)\nonumber\\&-&i\lambda\frac{k_{\mu}k_{\nu}}{k^{4}}
f(k^{2}),\end{eqnarray}

\noindent where $P_{\mu\nu}\equiv g_{\mu\nu}-{k_{\mu}k_{\nu}}/{k^{2}}$ and an
arbitrary function $f(k^{2})$ has been incorporated to the longitudinal part.
Accordingly, $\Lambda^\mu$ can be splited as follows

\begin{equation}\Lambda^\mu=\Lambda^{\mu}_{g}+\Lambda^{\mu}_{\epsilon}+
\Lambda^{\mu}_{L},\end{equation}

\noindent where the subscripts $g$, $\epsilon$, and $L$ make reference to those
pieces of $D_{\mu\nu}$ proportional to $g_{\mu\nu}$,
$\epsilon_{\mu\nu\rho}k^{\rho}$ and $k_{\mu}k_{\nu}$, respectively. The
computation of $\Lambda^{\mu}_{L}$ is straightforward and yields

\begin{equation}\Lambda^{\mu}_{L}=i\lambda\bar v^{(+)}({\bf p}_{1}^{\prime})
\gamma^{\mu} v^{(-)}({\bf p}_{1})\int\frac{d^{3}k}{(2\pi^{3})}\frac{f(k^{2})}
{k^2},\end{equation}

\noindent where, as in Ref.~\cite{Gi}, $v^{(-)}({\bf p}_{1})$ ($\bar
v^{(+)}({\bf p}_{1}^{\prime}$)) is a two-component spinor describing a free
electron of two-momentum ${\bf p}_{1}$ (${\bf p}_{1}^{\prime}$) in the initial
(final) state. Since $\Lambda^{\mu}_{L}$ does not depend upon the momentum
transfer $q\equiv p_{1}^{\prime}-p_{1}$, it can be absorbed into the
wave-function renormalization constant. When computing $\Lambda^{\mu}_{g}$ and
$\Lambda^{\mu}_{\epsilon}$ only zero- and first-order terms in ${q}/{m}$ will
be retained, since our interest is restricted to the non-relativistic regime.
Furthermore, all loop momentum integrals are ultraviolet-finite and there is,
then, no need for regularization. After absorbing the zero-order terms into
the wave-function renormalization constant, one arrives at

\begin{eqnarray}\label{at11}\Lambda^{\mu (1)}_{g}(q)=&&\frac{N_{g}}{16
\pi}\left(\frac{e^2}{m}\right)\epsilon^{\mu\nu\rho}\left (\frac
{q_{\nu}}{m}\right)\nonumber\\&\times&\bar v^{(+)}({\bf p}_{1}^{\prime})
\gamma_{\rho}v^{(-)}({\bf p}_{1}),\end{eqnarray}

\begin{eqnarray}\label{at12}\Lambda^{\mu(1)}_{\epsilon}(q)=
&-&\frac{N_{\epsilon}}{16\pi}\left(\frac{e^2}{m}\right)
\epsilon^{\mu\nu\rho}\left (\frac{q_{\nu}}{m}\right)\nonumber\\ &\times&\bar
v^{(+)}({\bf p}_{1}^{\prime})\gamma_{\rho}v^{(-)}({\bf p}_{1}),\end{eqnarray}

\noindent where

\begin{eqnarray}\label{at13}N_{g}&=&-3+3\frac{e^2}{8\pi m}\nonumber\\&&
+2\left(1-\frac{3e^{4}}{(16\pi m)^{2}}\right)\ln\left(1+\frac{16\pi
m}{e^2}\right),\end{eqnarray}

\noindent and

\begin{equation}\label{at14}N_{\epsilon}=-\frac{e^2}{\pi
m}\ln\left(1+\frac{16\pi m}{e^2}\right).\end{equation}

\noindent From Eqs.~(\ref{at11}--\ref{at14}) it follows that
$\Lambda^{\mu(1)}(q)$ behaves as $(e^2/m)\ln(e^2/m)$ when $e^2/m\rightarrow 0$
(${\theta_{in}}/{m}\rightarrow 0$), while power counting indicates that the
full vertex insertion $\Lambda^{\mu}(q)$ (Fig.~\ref{fi2}) diverges
logarithmically at the just mentioned infrared limit. The fact that
$\Lambda^{\mu (1)}(q)$ exhibits an improved infrared behavior is not a
peculiarity of the particular insertion under analysis but applies to any
vertex part involving an arbitrary number of exchanged massive vector
particles. Indeed, the leading infrared divergence of any of these parts only
shows up in those terms containing even powers of the momentum transfer $q$, as
can be seen by setting  to zero the loop momenta in the numerators of the
corresponding Feynman integrals. Thus, the terms linear in $q$ exhibit a milder
infrared behavior.

By using the technique described in \cite{Gi} one finds the correction $\Delta
V^{\rm QED_3}$ to the potential (\ref{at3}) arising from the diagrams in
Fig.~\ref{fi3},

\begin{eqnarray}\label{at15}e\Delta V^{\rm QED_3}&=&
\left(\frac{e^2}{8\pi}\right)
\left(\frac{e^2}{m}\right)(N_g-N_{\epsilon})\nonumber\\
&\times&\left[\frac{e^2}{16\pi^2m}\left (1+\frac{e^2}{8\pi
m}\right)K_0\left(\frac{e^2}{8\pi}r\right)\right.\nonumber\\
&&\;-\left.\frac{\delta^2({\bf r})}{m^2}+\frac{e^2}{8\pi^2m^2}K_1
\left(\frac{e^2}{8\pi}r\right)\frac{L}{r}\right].\end{eqnarray}

\noindent From (\ref{at13}--\ref{at15}) it follows that $e\Delta V^{\rm QED_3}$
also behaves as $(e^2/m)\ln(e^2/m)$ when $e^2/m\rightarrow 0$, and turns out to
be negligible if $e^2/m\ll 1$. This establishes the region of validity of our
results.

We turn next to the investigation of the existence of bound states of two
identical fermions of mass $m$ interacting through the non-relativistic
potential $V^{\rm QED_3}$ given by (\ref{at3}). The corresponding radial
Schr\"odinger equation is found to read

\begin{equation}{\cal H}_{l}R_{nl}(y)=\epsilon_{nl}R_{nl}(y),\end{equation}

\noindent where

\begin{equation}\label{at17}{\cal H}_{l}R_{nl}(y)=-\left(\frac{\partial^{2}
R_{nl}}{\partial y^{2}}+\frac{1}{y}\frac{\partial R_{nl}}{\partial y}
\right)+U^{eff}_{l}R_{nl},\end{equation}

\begin{eqnarray}\label{at18}U^{eff}_{l}(y)=&&\frac{l^{2}}{y^{2}}+\frac
{\alpha_{in}}{2}\left(1+\beta_{in}\right)K_{0}(y)\nonumber\\
&+&\frac{\alpha_{in}l}{y^2}\left[1-y K_{1}(y)\right],\end{eqnarray}

\noindent  $y=e^2r/(8\pi)$, $\alpha_{in}=e^2/(\pi|\theta_{in}|)=8$,
$\beta_{in}=m/|\theta_{in}|=8\pi m/e^2$, $\epsilon_{nl}={64\pi^2mE_{nl}}/{e^4}$
and $E_{nl}$ is the energy eigenvalue. A straightforward analysis of
(\ref{at18}) reveals that electron-electron bound states are only possible for
$l=-1,-3,-5$ and $-7$. Had we chosen the negative sign for $\bar m$ we
would have obtained $\theta_{in}=+e^2/(8\pi)$ and bound
states, with identical energy eigenvalues, for $l=+1,+3,+5$ and $+7$.

The existence of bound states for the potential given in (\ref{at18}) was
examined numerically by means of a stochastic variational algorithm. In this
way we were able to quickly identify the state of minimun energy ($n=0$) for a
given  $l$ and a variety of values of the parameter $\beta_{in}$.

In order to be able to represent the wave function $R_{0l}(y)$ and the
expectation value of the energy $\epsilon_{0l}$ numerically, we must chose a
discrete collection of points along the radial direction. Hence, we calculated

\begin{equation}\epsilon_{0l}=\Frac{\sum_{i=0}^{i_{max}}
\Delta y\;y_i\left[\left(\frac{\Delta
R_{0l}}{\Delta y}\right)_{i}^2
+U_{l,i}^{eff}R_{0l,i}^{2}\right]}{\sum_{i=0}^{i_{max}}\Delta
y\;y_i\;R_{0l,i}^{2}},\end{equation}

\noindent where the values of $\Delta y$ and $i_{max}$ were chosen to ensure a
big enough range of integration, and a small enough integration interval. We
then varied the wave function randomly, accepting any changes that decreased
the
energy \cite{f4}.
We found that, starting from an arbitrary initial guess for $R_{0l}$
it relaxed very fast towards a stable functional configuration.
The expectation value of the energy soon became negative,
confirming the existence of a bound state.  One of these
configurations, together with the corresponding potential, is displayed in
Fig.~(\ref{fi4}).

We also measured the expectation value of the radius in the
state thus obtained, which gives us an idea of its size.
 The stability of the results was tested against
variations of the
range of integration and the size of the integration interval.

In Table~\ref{tb1} we present our numerical results, calculated assuming
$  m $ to be the electron mass. The energy eigenvalues for  $ l =
-5$ and $ -7$ are very close to those correponding to $ l=-3$ and $ -1$,
respectively, and for this reason they have not been included in
Table~\ref{tb1}. The average radius of the
bound state is given in \AA, and the binding energy in $eV$. The corresponding
approximate dissociation temperature in $K$ is also given. The errors quoted
were
evaluated from the small variations observed in the results when the range and
interval of integration were varied.

It is interesting to note that there are ranges of the parameters $l$ and
$\beta_{in}$ where the results are numerically consistent with the observed
transition temperatures of high--T$_{c}$ superconductors.
We do not think that this coincidence is merely
accidental, although we are far from claiming  that QED$_3$ appropriately
describes all features of high--T$_c$ superconductivity.
Notice that if the initial state of the system only contains low energy
electrons, energy conservation forbids the production of electron--positron
pairs and, hence, only electron--electron bound states can be formed.
No electron--positron bound state can show up and, as a
consequence, only the electron--electron bound state may serve to characterize
the many body ground state. Of course, it remains to be shown that the theory
exhibits a phase transition and that the many body ground
state is a condensate of electron--electron pairs. However, if this turns out
to be the case we find reasonable to think that some of the dissociation
temperatures given in this paper should  at least be a rough approximation
for the true transition temperatures.

\acknowledgements

We acknowledge Prof. R. Jackiw for reading the manuscript and for many valuable
criticisms. One of us (AJS) thanks the Center for Theoretical Physics of MIT
for
the kind hospitality.

This work has been supported in part by Conselho Nacional de Desenvolvimento
Cient\'{\i}fico e Tecnol\'ogico (CNPq), Brazil, by Funda\c{c}\~ao de Amparo \`a
Pesquisa do Estado de S\~ao Paulo (FAPESP), Brazil, and by Coordena\c{c}\~ao de
Aperfei\c{c}oamento de Pessoal do Ensino Superior (CAPES), Brazil.

The computer work was realized mostly in the
workstations of the Department of Mathematical Physics, which were aquired
through grants from FAPESP/CNPq.

\newpage
\begin{center}
FIGURES
\end{center}
\figure{Wavy lines represent free photons, while dashed lines refer to massive
vector particles.\label{fi1}}

\figure{The vertex insertion $\Lambda^{\mu}$.\label{fi2}}

\figure{Diagrams contributing to the potential $\Delta V^{\rm QED_3}$.
\label{fi3}}

\figure{The  bound state wave function (dashed-line) and the potential
$U^{eff}_{l}(y)$ (continuous line)
for $l=-3$ and $\beta_{in}=3000$. The vertical scale refers only to the
potential. The normalization of the wave function is arbitrary.\label{fi4}}
\newpage
\begin{center}
TABLES
\end{center}

\begin{table}
\centering
\caption{The size and energy of the bound states.\label{tb1}}
\begin{tabular}{c|c|ccc}
 $l$  & $\beta_{in}$ & size (\AA) & energy ($10^{-3}eV$) & temperature ($K$) \\
\tableline
 $-1$ & $1000$ & $\;\:88\pm 2$
& $\;\:-4.8\;\:\pm 0.1\;\:$ & $   \;\:55.\;\:\pm 1.\;\:$ \\
      & $3000$ & $   300\pm 5$
& $\;\:-0.42   \pm 0.01   $ & $\;\:\;\:4.8   \pm 0.1   $ \\
\tableline
 $-3$ & $1000$ & $\;\:64\pm 1$
& $   -19.6\;\:\pm 0.4\;\:$ & $      228.\;\:\pm 4.\;\:$ \\
      & $3000$ & $   218\pm 4$
& $\;\:-1.7\;\:\pm 0.04   $ & $   \;\:19.7   \pm 0.4   $ \\
\end{tabular}
\end{table}


\begin{references}

\bibitem[*]{ast} On leave of absence from Instituto de F\'{\i}sica,
Universidade Federal do Rio Grande do Sul, Caixa Postal 15051, 91500 - Porto
Alegre, RS, Brazil.

\bibitem[\dag]{loa} On leave of absence from Instituto de F\'{\i}sica,
Universidade de S\~ao Paulo, Caixa Postal 20516, 01498 - S\~ao Paulo, SP,
Brazil.

\bibitem{Gi} H. O. Girotti, M. Gomes and A. J. da Silva, Phys Lett. {\bf B274},
357 (1992).

\bibitem{Ja} S. Deser, R. Jackiw and S. Templeton, Ann. Phys. (NY) {\bf 140},
372 (1982).

\bibitem{f1} In Eq.~(\ref{at1}), $e$ is the fermion charge, $\vec r$ is the
relative distance between electrons, $r=|\vec r|$, $L=\vec r\times\vec p$ is
the orbital angular momentum, whose eigenvalues are denoted by $l$, and $\vec
p$ is the relative linear momentum of the electrons. The fermion and
topological masses are $m$ and $\theta$, respectively, while $K_0$ and $K_1$
designate the modified Bessel functions. The linear dependence of $V$ on $L$
accounts for the breaking of parity and time reversal invariance in the
non-relativistic approximation.

\bibitem{f5} The potential (\ref{at1}) looks similar to the one derived by
Kogan \cite{Ko}
for the same problem. There are some important differences, however, that we
would like to stress. The derivation in Ref. \cite{Gi} is solely based on
relativistic quantum field theory \cite{f6}. On the other hand,
in Ref.\cite{Ko} the corresponding potential is determined in two steps.
First, the
$A_\mu$ vector created by a point charge is computed. Then, the quantum
mechanical Hamiltonian describing the low
energy relative motion of the two electrons is assumed to be that of a charged
particle in the presence of the $A_\mu$ external field. As a consequence,
an extra term proportional to
$\left[1-|\theta|r K_{1}\left(|\theta|r\right)\right]^{2}$ arises in the
formulation of Ref. \cite{Ko}.

\bibitem{f6} For a similar derivation of the effective electron-electron
potential arising from QED$_4$ see, for instance, J. J. Sakurai, "Advanced
Quantum Mechanics" (Addison--Wesley, Mass., 1967); A. I. Akhiezer and V. B.
Berestetskii, "Quantum Electrodynamics" ( John Wiley, N. Y., 1965).

\bibitem{Ko} Ya. I. Kogan, JETP Lett. {\bf 49}, 225 (1989).

\bibitem{Bl} F. Bloch and A. Nordsieck, Phys. Rev. {\bf 52}, 54 (1937).

\bibitem{Bo} N. N. Bogoliubov and D. V. Shirkov, ``Introduction to the Theory
of Quantized Fields'' (John Wiley, N. Y. , 1980).

\bibitem{Sz} A. V. Svidzinskiy, JETP {\bf 31}, 324 (1956).

\bibitem{f2} This does not constitute a serious limitation for QED$_4$, as far
as the infrared structure of the theory is concerned. In fact, in that case
these diagrams are, on the one hand, free of infrared singularities while,
on the
other hand, they do not give rise to a mechanism for dynamical mass generation.

\bibitem{Re} A. N. Redlich, Phys. Rev. Lett. {\bf 52}, 18 (1983); Phys. Rev.
{\bf D29}, 2366 (1984).

\bibitem{f3} Throughout this paper we use natural units $(c=\hbar=1)$. Our
metric is $g_{00}=-g_{11}=-g_{22}=1$, while for the $\gamma$ matrices we adopt
the representation $\gamma^{0}=\sigma_{3}$, $\gamma^{1}=i\sigma^{1}$,
$\gamma^{2}=i\sigma^{2}$, where $\sigma^{i}$, $i=1,2,3$ are the Pauli spin
matrices.

\bibitem {f7} For the vertex corrections in the MCS theory see
I. Kogan, Phys. Lett. {\bf B262}, 83 (1991); I. Kogan and G. W. Semenoff,
Nucl. Phys. {\bf B368}, 718 (1992).

\bibitem{f4} Hence, only the minimun energy eigenvalue ($n=0$), for a given $l$
, can be found through this method.
\end{references}
\end{document}